\documentclass[a4,12pt]{article}

\usepackage{a4}

\usepackage{amsmath}
\usepackage{amsfonts}
\usepackage{amssymb}
\usepackage{graphicx}
\usepackage{cite}

\def\tr{\mathop{\rm tr}\nolimits}

\newcommand{\CC}{\mathbb{C}}
\newcommand{\ZZ}{\mathbb{Z}}
\newcommand{\RR}{\mathbb{R}}

\usepackage{bm}

\newcommand{\wt}{\widetilde}
\newcommand{\ol}{\overline}

\def\Pexp{\mathop{\rm Pexp}\nolimits}

\begin{document}
%%%%%%%%%%%%%%%%%%%%%%%%%%%%%%%%%%%%%%%%%%%%%%%%%%%%%%%%%%%%%%%%
%titlepage
\begin{titlepage}
\title{
\vspace{-1.5cm}
\begin{flushright}
{\normalsize TIT/HEP-682\\ March 2021}
\end{flushright}
\vspace{1.5cm}
\LARGE{Flavor symmetries of six-dimensional ${\cal N}=(1,0)$
theories from AdS/CFT correspondence}}
\author{
Shota {\scshape Fujiwara\footnote{E-mail: s.fujiwara@th.phys.titech.ac.jp}},
Yosuke {\scshape Imamura\footnote{E-mail: imamura@phys.titech.ac.jp}},
and
Tatsuya {\scshape Mori\footnote{E-mail: t.mori@th.phys.titech.ac.jp}}
\\
\\
{\itshape Department of Physics, Tokyo Institute of Technology,} \\ {\itshape Tokyo 152-8551, Japan} }

\date{}
\maketitle
\thispagestyle{empty}

%abstract
\begin{abstract}
We calculate the superconformal indices of
a class of six-dimensional ${\cal N}=(1,0)$ superconformal field theories realized on M5-branes
at $\CC^2/\ZZ_k$ singularity
by using the method developed in previous works of the authors and collaborators.
We use the AdS/CFT correspondence, and finite $N$ corrections are included as the contribution of
M2-branes wrapped on two-cycles in $S^4/\ZZ_k$.
We confirm that the indices are consistent with the
expected flavor symmetries.
\end{abstract}
\end{titlepage}
\tableofcontents
%%%%%%%%%%%%%%%%%%%%%%%%%%%%%%%%%%%%%%%%%%%%%%%%%%%%%%%%%%%%%%%%
\section{Introduction}
Recent progress of quantum field theory
has enabled us quantitative analyses of strongly coupled field theories.
In particular, for supersymmetric theories realized in string/M-theories
we can use different powerful methods
like dualities and localization.
In this paper we will study a class of such theories:
six-dimensional ${\cal N}=(1,0)$ superconformal
field theories realized on
a stack of $N$ M5-branes placed at the fixed locus of
the $\CC^2/\ZZ_k$ orbifold.
A theory in this class has holographic description:
M-theory in $AdS_7\times S^4/\ZZ_k$.
A quiver gauge theory description is
also useful to describe the tensor branch of the theory
\cite{Brunner:1997gk,Hanany:1997sa,Brunner:1997gf,Hanany:1997gh},
and the root of the tensor branch,
where the superconformal symmetry restores,
corresponds to the strong gauge coupling limit.

An interesting property of such ${\cal N}=(1,0)$ theories
is non-trivial flavor symmetry depending on $k$ and $N$
\cite{Ohmori:2015pia,Hanany:2018vph,Bergman:2020bvi,Apruzzi:2020eqi,Bertolini:2015bwa,Merkx:2017jey,Bhardwaj:2020ruf}.
A subtle point is that the symmetry may be
different from what is read off from the
corresponding quiver gauge theory.
It was proposed that a certain discrete symmetry of
the quiver gauge theory, which is not manifest perturbatively,
is gauged in the strong coupling limit and as a result
the flavor symmetry is reduced \cite{Hanany:2018vph}.
For example, in the case of $N=k=2$, although the flavor symmetry of the quiver gauge theory is $SO(8)$,
that of the superconformal theory
is $SO(7)$ \cite{Ohmori:2015pia}.
See Table \ref{flavor.tbl} for the flavor symmetries for different $k$ and $N$.
\begin{table}[htb]
\caption{The flavor symmetries of interacting ${\cal N}=(1,0)$ theories.}\label{flavor.tbl}
\centering
\begin{tabular}{ccc}
\hline
\hline
& $N=2$ & $N\geq3$ \\
\hline
$k=2$ & $SO(7)$ & $SU(2)_a\times SU(2)_b\times SU(2)_F$ \\
$k\geq3$ & $SU(2k)$ & $SU(k)_a\times SU(k)_b\times U(1)_F$ \\
\hline
\end{tabular}
\end{table}
The purpose of this paper is to confirm such
flavor symmetries
on the AdS side.

For generic values of $k$ and $N$
the flavor symmetry is $SU(k)_a\times SU(k)_b\times U(1)_F$.
In addition, we also have $SU(2)_R$ symmetry.
These symmetries are manifest on the AdS side;
$U(1)_F\times SU(2)_R$ is the isometry of $S^4/\ZZ_k$, and two $SU(k)$ symmetries
are associated with the two $A_{k-1}$ singularities at the fixed points.
In the case of $k=2$ and $N\geq 3$, $U(1)_F$ is enhanced
to $SU(2)_F$, and this is also understood as the isometry of $S^4/\ZZ_2$.

The enhancement for $N=2$ is more interesting.
This symmetry enhancement is not manifest on the AdS side,
and it is interesting to study how this is realized.
We confirm this symmetry enhancement by
calculating
the superconformal index using the AdS/CFT correspondence.
This cannot be seen in the large $N$ limit, and
to confirm such symmetry enhancement
we need to include finite $N$ corrections.

To calculate the finite $N$ corrections to the superconformal index
we use the method developed in the previous work \cite{Arai:2020uwd}
of the authors and collaborators for $(2,0)$ theories.
See also \cite{Arai:2019xmp,Arai:2019wgv,Arai:2019aou,Arai:2020qaj} for similar analysis of the 4d superconformal index.
We calculate the corrections as a contribution from wrapped M2-branes.
We focus only on the single-wrapping contribution.
For a configuration with more than one wrapped M2-branes
its contribution is given in the form of integral over gauge fugacities.
We have not yet understood how to choose integration contours in the integrals,
and we leave the analysis of such contributions for future work.

This paper is organized as follows.
In the next section, we summarize basic properties of the theory we study.
In particular, we briefly explain the flavor symmetry.
In Section \ref{secmtheory}, we give a formula
we use to calculate the index.
In Section \ref{seccheck}
by using the formula we calculate the index for small $k$ and $N$.
For $N=1$ and arbitrary $k$ the ${\cal N}=(1,0)$ theory is free
and we can directly calculate the index without using the holographic description,
and we can confirm the formula gives the correct index.
This analysis is done for small $k$ in Section \ref{seccheckn1}.
Results for $N=2$ and $N=3$ are shown in subsection \ref{seccheckn2} and \ref{seccheckn3},
respectively, and the consistency with the flavor symmetries in Table \ref{flavor.tbl}
is confirmed.
The last section is devoted to discussions.

%%%%%%%%%%%%%%%%%%%%%%%%%%%%%%%%%%%%%%%%%%%%%%%%%%%%%%%%%%%%%%%%%%%%%%%%%
\section{6d $(1,0)$ theories}
\subsection{Setup}
An ${\cal N}=(1,0)$ theory we discuss is defined as the theory
on $N$ M5-branes placed at $A_{k-1}$ singularity.
We consider M-theory in the background $\RR^{1,5}\times\CC^2/\ZZ_k\times\RR_T$.
Let $X_\mu$ ($\mu=0,1,\ldots,5$), $z_i$ ($i=1,2$), and $x_5$ be the coordinates of
$\RR^{1,5}$, $\CC^2$, and $\RR_T$, respectively.
We also define $x_m$ ($m=1,2,3,4$) by
\begin{align}
z_1=x_1+i x_2,\quad z_2=x_3+i x_4.
\end{align}

Let $R_{ab}$ ($a,b=1,\ldots,5$) be the generators of the rotation group $SO(5)_R$ in the $x_a$ space.
We define the orbifold by $\ZZ_k$ generated by
\begin{align}
\exp\left(\frac{2 \pi i}{k}(R_{12}-R_{34})\right).
\label{zkgenerator}
\end{align}
This acts on $(z_1,z_2,x_5)$ as
\begin{align}
(z_1,z_2,x_{5})\rightarrow (e^{2\pi i/k}z_1,e^{-2\pi i/k}z_2,x_{5}).
\label{zkaction}
\end{align}

We put $N$ M5-branes at $x_1=\cdots=x_5=0$.
If it were not for the orbifolding the $A_{N-1}$-type
${\cal N}=(2,0)$ theory would be realized on
the worldvolume of the M5-branes.
The orbifolding breaks the $\mathcal{N}=(2,0)$ supersymmetry
down to  $\mathcal{N}=(1,0)$.
At the same time, the $SO(5)_R$ symmetry is broken to $SU(2)_R\times  U(1)_F$
for $k\geq3$.
$U(1)_F$ is replaced by $SU(2)_F$ for $k=2$.
The $SU(2)_R$ is the $R$-symmetry of the 6d $(1,0)$ SCFTs, while
$U(1)_F$ or $SU(2)_F$ does not act on the ${\cal N}=(1,0)$ supercharges and is treated as
a flavor symmetry.
In addition, the orbifold singularity provides $SU(k)$ flavor symmetry.
The singular locus $\RR^{1,5}\times\RR_T$ is devided by the
M5-branes at $x_5=0$ into two parts: the $x_5>0$ part and the $x_5<0$ part.
Correspondingly, we have two copies of $SU(k)$ symmetry which we denote by
$SU(k)_a$ and $SU(k)_b$.
In summary, the bosonic global symmetry is
\begin{align}
SO(2,6)_{\rm conf}\times SU(2)_R \times G_{\rm flavor},
\end{align}
where $SO(2,6)_{\rm conf}\times SU(2)_R$ is the bosonic subgroup of the 6d $\mathcal{N}=(1,0)$
superconformal symmetry $OSp(8|2)$ and
the flavor symmetry $G_{\rm flavor}$ is generically given by
\begin{align}
G_{\rm flavor}=U(1)_F\times SU(k)_a\times SU(k)_b.
\end{align}

We define the superconformal index as follows.
Let $H$ and $J_{ij}$ ($i,j=1,\ldots,6$) be the generators of $SO(2)_H\times SO(6)_{\rm spin}\subset SO(2,6)_{\rm conf}$,
and take $H$, $J_{12}$, $J_{34}$, and $J_{56}$ as Cartan generators.
$H$ is the Hamiltonian and $J_{ij}$ are spins.
We also take $R_{12}$ and $R_{34}$ as $SO(5)_R$ Cartan generators.
To define the index we need to choose one component of the supercharge.
We take the component with the following quantum numbers:
\begin{align}
{\cal Q} : (H,J_{12},J_{34},J_{56} ; R_{12},R_{34})=(+\tfrac{1}{2},-\tfrac{1}{2},-\tfrac{1}{2},-\tfrac{1}{2} ; +\tfrac{1}{2}, +\tfrac{1}{2}).
\end{align}
Note that
${\cal Q}$ is invariant under the orbifold group $\ZZ_k$ generated by (\ref{zkgenerator}).
The anticommutation relation between ${\cal Q}$ and its hermitian conjugate ${\cal Q}^\dagger$ is
\begin{align}
\Delta\equiv\{ {\cal Q},{\cal Q}^\dagger\}
=H-( J_{12}+ J_{34}+ J_{56})-2( R_{12}+ R_{34}).
\label{qqdagger}
\end{align}
Then we define the superconformal index by
\begin{align}
{\cal I}( q, y_a, u,a_i,b_i)
=\tr\left[(-1)^F x^{\Delta}
 q^{ H+\frac{1}{3}( J_{12}+ J_{34}+ J_{56})}
 y_1^{ J_{12}}
 y_2^{ J_{34}}
 y_3^{ J_{56}}
 u^{ R_{12}- R_{34}}
 \prod_{i=1}^{k-1} a_i^{F_{a,i}} b_i^{F_{b,i}}
\right].
\end{align}
Due to the boson/fermion cancellation only the BPS operators with $\Delta =0$ contribute to the index,
and hence the index does not depends on $x$.
The choice of ${\cal Q}$ breaks the $SO(6)_{\rm spin}$ to $U(1)_{\rm spin}\times SU(3)_{\rm spin}$,
and
$y_1$, $y_2$, and $y_3$ are the $SU(3)_{\rm spin}$ fugacities constrained by $y_1y_2y_3=1$.
$F_{a,i}$ and $F_{b,i}$ are respectively Cartan generators of $SU(k)_a$ and $SU(k)_b$.
$u$ is the fugacity for $U(1)_F$ generated by $R_{12}- R_{34}$.
Note that for $k=1$ this index agree with the ${\cal N}=(2,0)$ superconformal index in \cite{Arai:2020uwd}.

%%%%%%%%%%%%%%%%%%%%%%%%%%%%%%%%%%%%%%%%%%%%%%%%%%%%%%%%%%%%%%%%
\subsection{Flavor symmetries}
For the analysis of the operator spectrum and the flavor symmetry of the theory
it is convenient to consider the quiver gauge theories realized in the tensor branch.
By taking $U(1)_F$ orbits as M-theory circles we can regard the system as a
type IIA brane configuration.
$N$ M5-branes become $N$ NS5-branes,
and the $A_{k-1}$ singularity becomes a stack of $k$ D6-branes.

The $(2,0)$ tensor multiplet on an M5-brane separate into a $(1,0)$ tensor multiplet and a $(1,0)$ hypermultiplet
on the corresponding NS5-brane,
and the scalar component in the $(1,0)$ tensor multiplet corresponds to the location of the NS5-branes in the $x_5$ direction.
At a generic point in the tensor branch
all the NS5-branes are separated one by one in the $x_5$ direction,
and a linear quiver gauge theory is realized on the D6-branes.

\begin{figure}[htb]
\begin{center}
\includegraphics[scale=1]{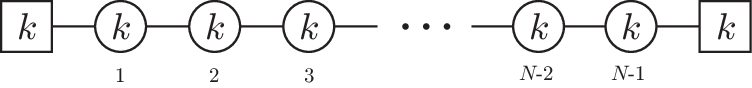}
\caption{The linear quiver diagram of the gauge theory realized in the tensor branch is shown.}
\label{linearquiver_gene}
\end{center}
\end{figure}

The worldvolume of the stack of D6-branes is divided into $N+1$ parts by the NS5-branes.
We label the NS5-branes by $i=1,2,\ldots,N$.
The D6-branes suspended between two NS5-branes $i$ and $i+1$ give
$SU(k)_i$ gauge group
while two semi-infinite parts of D6-branes give the flavor symmetries $SU(k)_a\equiv SU(k)_0$
and $SU(k)_b\equiv SU(k)_N$.
Let $(h_i,\wt h_i)$ be the hypermultiplet arising from open strings crossing
the $i$-th NS5-brane.
$h_i$ and $\wt h_i$ belong to the bi-fundamental representations $(k,\ol k)$ and $(\ol k,k)$,
respectively, of $SU(k)_{i-1}\times SU(k)_i$.
The $SU(N)$ groups and the hypermultiplets are depicted as the linear quiver diagram in Figure \ref{linearquiver_gene}.
In addition, we also have degrees of freedom that are implicit in the diagram;
in each gauge node there exists a tensor multiplet corresponding to the degrees of freedom of the NS5-brane.

%%%%%%%%%%%%%%%%%%%%%%%%%%%%%%%%%%%%%%%%%%%%%%

Among different gauge invariant operators
let us focus on two classes of operators.
The first class includes operators defined by
\begin{align}
S_{ij}=(h_0)_{ia}(\wt h_0)_{aj},\quad
S'_{ij}=(\wt h_N)_{ia}(h_N)_{aj}.
\end{align}
The other class includes
\begin{align}
L_{ij}=(h_0)_{ia_1}(h_1)_{a_1a_2}\cdots(h_N)_{a_Nj},\quad
L'_{ij}=(\wt h_N)_{ia_N}\cdots(h_1)_{a_2a_1}(h_0)_{a_1j}.
\end{align}
These operators play an important role when we discuss the flavor symmetry.

The operators $S_{ij}$ and $S'_{ij}$ belong to the adjoint representations
of $SU(k)_0$ and $SU(k)_N$,
respectively.
They have dimension $4$
and are the primary operators of the current multiplets of the
generic flavor symmetry
\begin{align}
G_{\rm flavor}=SU(k)_0\times SU(k)_N\times U(1)_F.
\label{gensym}
\end{align}
(Although there are $N$ classical $U(1)$ symmetries only
one of them is anomaly free.)

The operators $L_{ij}$ and $L'_{ij}$,
which belong to the bi-fundamental representations $(k,\ol k)_{+1}$ and $(\ol k,k)_{-1}$
of $G_{\rm flavor}$ in (\ref{gensym}),
respectively, have dimension $2N$.
These operators appear in the spectrum only when $N$ is finite,
and play a role similar to
baryonic operators in four-dimensional quiver gauge theories.
They are expected to correspond to wrapped M2-branes on the gravity side.
Indeed, their dimension coincide with the mass of an M2-brane wrapped
around a large $S^2/\ZZ_k$
in the unit of the inverse AdS radius.

The flavor symmetry (\ref{gensym}) is enhanced if one of $k$ or $N$
becomes $2$.
If $k=2$ and $N\geq3$ $U(1)_F$ is enhanced to $SU(2)_F$.
Correspondingly, the index is written in terms of $SU(2)_F$ characters.
This symmetry is manifest on the gravity side
as the isometry of $S^4/\ZZ_2$.

The enhancement for $N=2$ is more interesting.
If $N=2$ the operators $L$ and $L'$ have dimension $4$ as well as $S$ and $S'$,
and they give additional current multiplets.
As the result the flavor symmetry (\ref{gensym})
is enhanced to $SU(2k)$ for $k\geq 3$.
On the gravity side, this enhancement should be realized
when we include the contribution of wrapped M2-branes.

The $k=N=2$ case is most interesting.
In this case there are eight $SU(2)$ gauge symmetry doublet
in the hypermultiplets,
and we can write down $28$ gauge invariant dimension $4$ operators
forming the $SO(8)$ adjoint representation.
They correspond to the $SO(8)$ global symmetry of the
quiver gauge theory.
However, it is known that the symmetry is reduced to $SO(7)$
in a highly non-trivial way \cite{Ohmori:2015pia},
and it would be nice if we
can reproduce this flavor symmetry on the gravity side
by the index calculation.

%%%%%%%%%%%%%%%%%%%%%%%%%%%%%%%%%%%%%%%%%%%%%%%%%%%%%%%%%%%%%%%%%%%%%%
\section{Indices from M-theory in $AdS_7\times \bm{S}^4/\ZZ_k$}\label{secmtheory}
%%%%%%%%%%%%%%%%%%%%%%%%%%%%%%%%%%%%%%%%%%%%%%%%%%%%%%%%%
\subsection{Conjectural formula}

Based on the idea explained in detail in \cite{Arai:2020uwd}
we propose the formula
of the index for 6d $(1,0)$ theories
\begin{align}
\mathcal{I}^{(1,0)}_{N,k}
=\mathcal{I}^{\mathrm{bulk}}
\sum_{n_1,n_2=0}^\infty\mathcal{I}_{(n_1,n_2)}^{\mathrm{M2}}.
\label{idformula0}
\end{align}
This formula gives the index as the combination of contributions from
objects in the dual geometry
$AdS_7\times S^4/\ZZ_k$, where
the internal space $\bm{S}^4/\ZZ_k$ is defined by
\begin{align}
|z_1|^2+|z_2|^2+x_5^2=1,
\end{align}
together with the identification by the $\ZZ_k$ action (\ref{zkaction}).
${\cal I}^{\rm bulk}$ is the contribution of
Kaluza-Klein modes in the bulk.
We also include in ${\cal I}^{\rm bulk}$ the contribution from
localized modes at the fixed points of the orbifold.
${\cal I}^{\rm M2}_{(n_1,n_2)}$ are contributions of wrapped M2-branes in the internal space.
$n_1$ and $n_2$ are numbers of M2-brane wrapped on the two specific two-cycles $z_1=0$ and $z_2=0$, respectively.
As we will explain in \ref{wrapped.sec} the $q$ expansion of ${\cal I}_{(n_1,n_2)}$ starts from order $q^{2(n_1+n_2)N}$
terms.
In the large $N$ limit all contributions but ${\cal I}_{(0,0)}=1$ decouple
and the formula reduces to
$\mathcal{I}^{(1,0)}_{N=\infty,k}
=\mathcal{I}^{\mathrm{bulk}}$.
On the other hand, if $N$ is finite, all sectors labelled by $(n_1,n_2)$ contribute to the index.
${\cal I}_{(n_1,n_2)}$ for each $(n_1,n_2)$ is calculated as the index of the theory realized on the
wrapped M2-branes by the standard localization formula.
If $n_1+n_2\geq2$ the formula includes non-trivial gauge integrals, and unfortunately we have not yet found
systematic rules for the integration contours.
For this reason we leave the analysis of $n_1+n_2\geq2$ for future work and
in this paper we focus only on the single-wrapping sectors $(n_1,n_2)=(1,0)$ and $(0,1)$.
Namely, we consider the formula
\begin{align}
\mathcal{I}^{(1,0)}_{N,k}
=\mathcal{I}^{\rm grav}_{N,k}
+{\cal O}(q^{4N}),\quad
\label{idformula2}
\end{align}
where ${\cal I}^{\rm grav}_{N,k}$ is defined by
\begin{align}
\mathcal{I}^{\rm grav}_{N,k}
=\mathcal{I}^{\mathrm{bulk}}
\left(1+\mathcal{I}_{(1,0)}^{\mathrm{M2}}+\mathcal{I}_{(0,1)}^{\mathrm{M2}}\right).
\label{idformula}
\end{align}
With the conjectural formula (\ref{idformula2}),
we can calculate the index for an arbitrary $N$ and $k$ up to the expected error of order $q^{4N}$.

%%%%%%%%%%%%%%%%%%%%%%%%%%%%%%%%%%%%%%%%%%%
\subsection{The bulk contribution}
Let us first consider ${\cal I}^{\rm bulk}$, which gives the large $N$ index.
This is given by the plethystic exponential of the single-particle index,
which is the sum of two contributions:
the supergravity Kaluza-Klein modes in the internal space
$\bm{S}^4/\ZZ_k$ and the vector multiplets localized at
the two fixed points of $\bm{S}^4/\ZZ_k$.

The contribution of the Kaluza-Klein modes in $AdS_7\times \bm{S}^4$
without orbifolding
has already been studied in \cite{Bhattacharya:2008zy} and is given by
$\Pexp i_{\rm KK}$
with the single-particle index
\begin{align}
i_{\rm KK}
&=\frac{
 q^2\chi_{[1]}^u
- q^{\frac{8}{3}}\chi_{[0,1]}^y
+ q^{\frac{16}{3}}\chi_{[1,0]}^y
- q^6\chi_{[1]}^u
}{(1- u q^2)(1- u^{-1} q^2)
(1- y_1 q^{\frac{4}{3}})
(1- y_2 q^{\frac{4}{3}})
(1- y_3 q^{\frac{4}{3}})},
\end{align}
where $\chi_{[n]}^u$ are the $SU(2)$ characters defined by
\begin{align}
\chi_{[n]}^u=\frac{u^{n+1}-u^{-(n+1)}}{u-u^{-1}},   
\end{align}
and $\chi_{[m_1,m_2]}^y$ are the $SU(3)$ characters of the representations with Dynkin labels $[m_1,m_2]$.
The Kaluza-Klein modes in the orbifold $S^4/\ZZ_k$
is obtained by picking up the $\ZZ_k$ invariant modes
from the modes in $S^4$ \cite{Ahn:1998pb}.
Correspondingly, the single-particle index for the orbifold
is given by ${\cal P}_k i_{\rm KK}$,
where ${\cal P}_k$ is the projection operator associated with the $\ZZ_k$ orbifold
which acts on a function of the fugacity $u$ as
\begin{align}
\label{projop}
{\cal P}_k f(u)= \frac{1}{k} \sum_{l=0}^{k-1}f(e^{2\pi i l/k} u).
\end{align}

The other contribution we need to include in the single-particle index comes from
two $A_{k-1}$ singularities on $\CC^2/\ZZ_k$ at $(z_1,z_2,x_5)=(0,0,\pm1)$, where
the 7d $SU(k)_a\times SU(k)_b$ vector multiplets are localized.
In general, a gauge field in the bulk of AdS corresponds to a flavor symmetry on the boundary,
and the corresponding current multiplet
contributes to the index.
The corresponding single-particle index is
$i_F(\chi_{\mathrm{adj.}}^a+\chi_{\mathrm{adj.}}^b)$,
where $\chi^{a/b}_{\mathrm{adj.}}$ are characters of adjoint representations of the global $SU(k)_{a/b}$ symmetries
and $i_F$ is given by
\begin{align}
i_F=\frac{q^4}{(1-q^{\frac{4}{3}}y_1)(1-q^{\frac{4}{3}}y_2)(1-q^{\frac{4}{3}}y_3)}.
\label{iflavor}
\end{align}
Note that $i_F$ is independent of $u$ and we do not have to perform the $\ZZ_k$ projection.

By combining two contributions, we can calculate the index for the large $N$ limit.
For example, for $k=2$ we obtain
\begin{align}
\Pexp({\cal P}_2i_{\rm KK}+i_F(\chi^a_{[2]}+\chi^b_{[2]}))
&=1
-\chi_{[0,1]}^yq^{\frac{8}{3}}
+(\chi^u_{[2]}+\chi^a_{[2]}+\chi^b_{[2]}-\chi_{[1,1]}^y)q^4
\nonumber\\&
+((2+\chi^u_{[2]}+\chi^a_{[2]}+\chi^b_{[2]})\chi_{[1,0]}^y-\chi_{[2,1]}^y)q^{\frac{16}{3}}
+{\cal O}(q^{\frac{20}{3}}).
\end{align}
This includes the contribution of the ``center of mass'' free tensor multiplet.
The existence of such a decoupled free sector is suggested
by the coefficient ``$2$'' of
the term $\chi_{[1,0]}^yq^{\frac{16}{3}}$ in the above expansion,
which is identified as the contribution of two copies of
stress-tensor multiplets.
Such a free tensor multiplet exists for all $k$ and $N$, and we always remove its contribution
in the following calculation.
Namely, we define ${\cal I}^{\rm bulk}$ in (\ref{idformula2}) by
\begin{align}
{\cal I}^{\rm bulk}&=
\Pexp({\cal P}_ki_{KK}+i_F(\chi^a_{\rm adj}+\chi^b_{\rm adj})-i_{\rm tensor}),
\label{ninft}
\end{align}
where the single-particle index $i_{\rm tensor}$ of the free tensor multiplet is given by \cite{Bhattacharya:2008zy}
\begin{align}
i_{\mathrm{tensor}}&=\frac{-q^{\frac{8}{3}}\chi^y_{[0,1]}+q^4}{(1-q^{\frac{4}{3}}y_1)(1-q^{\frac{4}{3}}y_2)(1-q^{\frac{4}{3}}y_3)}.
\end{align}
For $k=2,3$ (\ref{ninft}) gives
\begin{align}
{\cal I}^{\rm bulk}_{k=2}&=
\Pexp({\cal P}_2i_{KK}+i_F(\chi^a_{[2]}+\chi^b_{[2]})-i_{\rm tensor})
\nonumber\\
&=1
+(\chi^u_{[2]}+\chi^a_{[2]}+\chi^b_{[2]})q^4
+(1+\chi^u_{[2]}+\chi^a_{[2]}+\chi^b_{[2]})\chi_{[1,0]}^yq^{\frac{16}{3}}
+{\cal O}(q^{\frac{20}{3}}),\\
%%%%%%%%
{\cal I}^{\rm bulk}_{k=3}&=
\Pexp({\cal P}_3i_{KK}+i_F(\chi_{[1,1]}^a+\chi_{[1,1]}^b)-i_{\rm tensor})
\nonumber\\
&=1
+(1+\chi_{[1,1]}^a+\chi_{[1,1]}^b)q^4
+(2+\chi_{[1,1]}^a+\chi_{[1,1]}^b)\chi_{[1,0]}^yq^{\frac{16}{3}}
+(-\chi_{[1]}^u+\chi_{[3]}^u)q^6
+{\cal O}(q^{\frac{20}{3}}).
\end{align}
These are interpreted as the indices in the large $N$ limit.
The $q^4$ terms in each index is the
contribution of the flavor current multiplets,
and we can read off the expected flavor symmetries $SU(2)^3$ for $k=2$ and $SU(3)^2\times U(1)$ for $k=3$.
We also confirm that all other terms are consistent with these flavor symmetries.

%%%%%%%%%%%%%%%%%%%%%%%%%%%%%%%%%%%%%%%%%%%%%%%%%
\subsection{Wrapped M2-branes on $\bm{S}^4/\ZZ_k$}\label{wrapped.sec}
Next we consider the contribution of M2-branes.
The worldvolume of a BPS M2-brane
is described by the intersection of $\bm{S}^4$ and a holomorphic surface
\cite{Mikhailov:2000ya,Bhattacharyya:2007sa}
\begin{align}
f(z_1,z_2)=0.
\end{align}
The consistency with the $\ZZ_k$ orbifolding require the function $f$ to satisfy
\begin{align}
f(e^{2\pi i/k}z_1,e^{-2\pi i/k}z_2)
=e^{2\pi iw/k}f(z_1,z_2),
\end{align}
where $w\in\ZZ/k\ZZ$ is the topological wrapping number.

In \cite{Arai:2020uwd}, in which the system without $\ZZ_k$ orbifolding
was studied, it was proposed that we can take only M2-brane configurations
given by monomials of the form $f(z_1,z_2)=z_1^{n_1}z_2^{n_2}$ and were shown that
the formula passes some non-trivial checks.
Let us adopt the same assumption.
The function $f(z_1,z_2)=z_1^{n_1}z_2^{n_2}$ gives the system with $n_1$ M2-branes
wrapped on $z_1=0$ and $n_2$ M2-branes wrapped on $z_2=0$.
The total topological wrapping number is $w=n_1-n_2$ mod $k$.
${\cal I}^{\rm M2}_{(n_1,n_2)}$ in (\ref{idformula0})
is the contribution from the specific wrapping sector with $(n_1,n_2)$.
We focus on the two sectors $(1,0)$ and $(0,1)$.
In the absence of the $\ZZ_k$
orbifolding, the contribution of the $(1,0)$ sector, a single M2-brane wrapped on $z_1=0$,
is \cite{Arai:2020uwd}
\begin{align}
\label{Im2z1}
(q^2 u)^N\Pexp i^{\mathrm{M2}}_{z_1=0},
\end{align}
with the single-particle index
$i^{\mathrm{M2}}_{z_1=0}$ given by
\begin{align}
i^{\rm M2}_{z_1=0}
&=\frac{q^{-2} u^{-1}- q^{\frac{2}{3}} u^{-1}\chi_{[0,1]}^y
+ q^{\frac{4}{3}}\chi_{[1,0]}^y- q^4}{1- q^2 u^{-1}}.
\end{align}
To obtain the index for the $\ZZ_k$ orbifold, we need two modifications.
First, we perform the $\ZZ_k$ projection on the single-particle index.
Second, we insert the character of the $SU(k)_a\times SU(k)_b$ bi-fundamental representation
because the wrapped M2-brane couples to the localized vector multiplets at the fixed points $(z_1,z_2,x_5)=(0,0,\pm1)$.
As the result, the contribution of the $(1,0)$ sector is given by
\begin{align}
\mathcal{I}^{\mathrm{M2}}_{(1,0)}&=(q^2 u)^N\chi^a_{\rm fund.} \chi^b_{\rm \overline{fund.}}\Pexp\left[\mathcal{P}_k i^{\mathrm{M2}}_{z_1=0}\right],
\label{i10}
\end{align}
where $\chi^{a/b}_{\rm fund.}$ ($\chi^{a/b}_{\rm \overline{fund.}}$) 
are characters of (anti-) fundamental representations of the $SU(k)_{a/b}$ symmetries.
The contribution of the other sector $(0,1)$ is
given from (\ref{i10}) by the replacement
$u \rightarrow u^{-1}$, $a_i\rightarrow a_i^{-1}$, and $b_i\rightarrow b_i^{-1}$.
$u\rightarrow u^{-1}$ is the Weyl reflection of $SO(5)_R$ exchanging $z_1$ and $z_2$.
The inversion of $a_i$ and $b_i$ are necessary because the cycle $z_2=0$
has the opposite topological wrapping number to the cycle $z_1=0$;
the former has $w=+1$ while the latter has $w=-1$.
After the replacement we obtain
\begin{align}
\mathcal{I}^{\mathrm{M2}}_{(0,1)}&=(q^2u^{-1})^N\chi^a_{\rm \overline{fund.}} \chi^b_{\rm fund.}\Pexp\left[\mathcal{P}_k i^{\mathrm{M2}}_{z_2=0}\right],
\label{i01}
\end{align}
where $i^{\mathrm{M2}}_{z_2=0}=i^{\mathrm{M2}}_{z_1=0}|_{u\rightarrow u^{-1}}$.

If $k=2$  the flavor characters appearing in (\ref{i10})
and (\ref{i01}) are the same, $\chi^a_{\rm fund.} \chi^b_{\rm \overline{fund.}}=\chi^a_{\rm \overline{fund.}} \chi^b_{\rm fund.}$, and
$\mathcal{I}^{\mathrm{M2}}_{(1,0)}+\mathcal{I}^{\mathrm{M2}}_{(0,1)}$
is invariant under the $SU(2)_R$ Weyl reflection $u\rightarrow u^{-1}$.
This is consistent with the symmetry enhancement
$U(1)_F\rightarrow SU(2)_F$.

%%%%%%%%%%%%%%%%%%%%%%%%%%%%%%%%%%%%%%%%%%%%%%%%%%%%%%%%%%%%%%%%
\section{Results and consistency check}\label{seccheck}
Now we are ready to calculate the index of the 6d $(1,0)$ theories by using our formula (\ref{idformula})
for different values of $k$ and $N$.

%%%%%%%%%%%%%%%%%%%%%%%%%%%%%%%%%%%%%%%%%%%%%%%%%
\subsection{Results for $N=1$}\label{seccheckn1}
The theory with $N=1$ is the free theory consisting of
the ``center-of-mass'' tensor multiplet and hypermultiplets belonging to
the bi-fundamental representation of $SU(k)_a\times SU(k)_b$.
The index with the tensor multiplet contribution removed
is given by
\begin{align}
\mathcal{I}^{(1,0)}_{N=1}=\Pexp\left[i_{\mathrm{hyper}}(\chi^a_{\rm fund.} \chi^b_{\rm \overline{fund.}} u+\chi^a_{\rm \overline{fund.}} \chi^b_{\rm fund.} u^{-1})\right],
\label{6dfree}
\end{align}
where $i_{\mathrm{hyper}}$
is given by \cite{Bhattacharya:2008zy}
\begin{align}
i_{\mathrm{hyper}}&=\frac{q^2}{(1-q^{\frac{4}{3}}y_1)(1-q^{\frac{4}{3}}y_2)(1-q^{\frac{4}{3}}y_3)}.
\end{align}

Let us compare the  index on the gravity side based on the formula (\ref{idformula})
with the free theory result from (\ref{6dfree}).
As we do not include the multiple-wrapping M2-branes,
the errors should start at $q^4$ terms and we check the agreement up to
errors of this order.

\subsubsection{$k=2$}

On the gravity side (\ref{idformula}) yields
\begin{align}
\mathcal{I}_{N=1,k=2}^{\mathrm{grav}}
&=1+ \chi ^{a}_{[1]} \chi ^{b}_{[1]} \chi ^{u}_{[1]} q^2
+ \chi ^{a}_{[1]} \chi ^{b}_{[1]} \chi ^{u}_{[1]} \chi_{[1,0]}^{y} q^{\frac{10}{3}}
+(\chi ^{a}_{[2]} +\chi ^{b}_{[2]} +\chi ^{u}_{[2]} )q^4+\mathcal{O}(q^{\frac{14}{3}}).
\end{align}
Expanding (\ref{6dfree}) with $k=2$ we obtain
\begin{align}
\mathcal{I}^{(1,0)}_{N=1,k=2}
&=\mathcal{I}_{N=1,k=2}^{\mathrm{gr}}
+\chi ^{a}_{[2]}  \chi ^{b}_{[2]}  \chi^u_{[2]}q^4+\mathcal{O}(q^{\frac{14}{3}}).
\end{align}
We can see the agreement up to the error of ${\cal O}(q^4)$
as expected.

\subsubsection{$k=3$}
For $k=3$ the result on the gravity side is
\begin{align}
\mathcal{I}_{N=1,k=3}^{\mathrm{grav}}&=
1+\left(u \chi _{[1,0]}^{a} \chi _{[0,1]}^{b}+u^{-1}\chi _{[0,1]}^{a} \chi _{[1,0]}^{b}\right)q^2
+\left(u \chi _{[1,0]}^{a} \chi _{[0,1]}^{b}\chi _{[1,0]}^{y} +u^{-1}\chi_{[0,1]}^{a} \chi _{[1,0]}^{b}\chi _{[1,0]}^{y}\right)q^{\frac{10}{3}}
\nonumber\\&
+\left(
1+\chi _{[1,1]}^{a}
+\chi _{[1,1]}^{b}
+u^2 \chi _{[0,1]}^{a} \chi _{[1,0]}^{b}
+u^{-2}\chi _{[1,0]}^{a} \chi_{[0,1]}^{b}
\right)q^4 +\mathcal{O}(q^{\frac{14}{3}}).
\end{align}
On the CFT side we obtain
\begin{align}
\mathcal{I}^{(1,0)}_{N=1,k=3}
&=
\mathcal{I}_{N=1,k=3}^{\mathrm{grav}}
+ \Bigg(
u^2 \chi _{[2,0]}^{a} \chi _{[0,2]}^{b}
+u^{-2}\chi _{[0,2]}^{a} \chi _{[2,0]}^{b}
+\chi _{[1,1]}^{a} \chi _{[1,1]}^{b}
\Bigg)q^4
+\mathcal{O}(q^{\frac{14}{3}}).
\end{align}
Again we can find agreement up to expected error terms of order $q^4$.

%%%%%%%%%%%%%%%%%%%%%%%%%%%%%%%%%%%%%%%%%%%%%%%%%
\subsection{Results for $N=2$}\label{seccheckn2}
When $N=2$, the generic flavor symmetry
$SU(k)_a\times SU(k)_b\times U(1)$ is enhanced to
$SU(2k)$ for $k\geq 3$ and $SO(7)$ for $k=2$ \cite{Ohmori:2015pia,Bah:2017gph,Hanany:2018vph}.
Then, the indices should be written in terms of the characters of the enhanced symmetries.
Let us confirm this for $k=2$ and $k=3$.
The expected errors due to double-wrapping configurations are of order $q^8$,
and we show the results below the order.

%%%%%%%%%%%%%%%%%%%%%%%%%%%%%%%%%%%%%%%%%%%%%%%%%
\subsubsection{$k=2$}
We consider the $k=N=2$ case first.
In this case the index should be written in terms of the $SO(7)$ character $\chi^{SO(7)}_{[l_1,l_2;l_3]}$.
The last component of the Dynkin labels corresponds to the
short root.
The formula (\ref{idformula2}) gives
\begin{align}
{\cal I}^{\rm grav}_{N=k=2}
&=1
+\chi^{SO(7)}_{[0,1;0]}q^4
+(1+\chi^{SO(7)}_{[0,1;0]})\chi^{y}_{[1,0]}q^{\frac{16}{3}}
\nonumber\\&
+\left((1+\chi^{SO(7)}_{[0,1;0]})\chi^{y}_{[2,0]}+
(1-\chi^{SO(7)}_{[1,0;0]})\chi^{y}_{[0,1]}\right)q^{\frac{20}{3}}
+{\cal O}(q^8).
\end{align}
This is correctly expanded in terms of $SO(7)$ characters.
We also confirm that it is not written in terms of characters of $SO(8)$, the symmetry of
the corresponding quiver gauge theory.

%%%%%%%%%%%%%%%%%%%%%%%%%%%%%%%%%%%%%%%%%%%%%%%%%
\subsubsection{$k=3$}
For $k=3$ and $N=2$
the expected flavor symmetry is $SU(6)$.
The formula (\ref{idformula2}) gives
\begin{align}
\mathcal{I}_{N=2,k=3}^{\mathrm{grav}}
&=1+\chi^{SU(6)}_{[1,0,0,0,1]}q^4
+(1+\chi^{SU(6)}_{[1,0,0,0,1]})\chi_{[1,0]}^{y}q^{\frac{16}{3}}
+\chi^{SU(6)}_{[0,0,1,0,0]}q^6
\nonumber\\&
+\left(1+\chi_{[1,0,0,0,1]}^{SU(6)}\right)\chi_{[2,0]}^{y}q^{\frac{20}{3}}
+\chi^{SU(6)}_{[0,0,1,0,0]}\chi_{[1,0]}^yq^{\frac{22}{3}}
+\mathcal{O}(q^{8}),
\end{align}
and this is correctly written in terms of $SU(6)$ characters.

%%%%%%%%%%%%%%%%%%%%%%%%%%%%%%%%%%%%%%%%%%%%%%%%%%%%
\subsection{Results for $N=3$}\label{seccheckn3}
In this subsection we show the results for $N=3$ calculated on the gravity side.
Because we do not have results we can compare,
we give the results simply as predictions.
The expected errors are of order $q^{12}$,
and we show the results below the order.

%%%%%%%%%%%%%%%%%%%%%%%%%%%%%%%%%%%%%%%%%%%%%%%%%%%%
\subsubsection{$k=2$}
The global symmetry for $N=3$ and $k=2$ is
$G_{\rm flavor}=SU(2)_a\times SU(2)_b\times SU(2)_F$.
The formula (\ref{idformula2}) gives
\begin{align}
\mathcal{I}_{N=3,k=2}^{\mathrm{grav}}
&=
1+(\chi ^{a}_{[2]} +\chi ^{b}_{[2]} +\chi ^{u}_{[2]} ) q^4
+(\chi ^{a}_{[2]}  \chi^y_{[1,0]}+\chi ^{b}_{[2]}  \chi^y_{[1,0]}+\chi ^{u}_{[2]}  \chi^y_{[1,0]}+\chi^y_{[1,0]}) q^{\frac{16}{3}}
\nonumber\\&
+\chi ^{a}_{[1]}  \chi ^{b}_{[1]}  \chi ^{u}_{[3]}  q^6
+(\chi^y_{[0,1]}+\chi ^{a}_{[2]}  \chi^y_{[2,0]}+\chi ^{b}_{[2]} 
   \chi^y_{[2,0]}+\chi^y_{[2,0]}+\chi ^{u}_{[2]}  (\chi^y_{[2,0]}-\chi^y_{[0,1]})) q^{\frac{20}{3}}
\nonumber\\&
+\chi ^{a}_{[1]}  \chi ^{b}_{[1]}  \chi ^{u}_{[3]}  \chi^y_{[1,0]} q^{\frac{22}{3}}
+(\chi ^{a}_{[4]} +\chi ^{a}_{[2]}  \chi ^{b}_{[2]} +\chi ^{b}_{[4]} +2 \chi ^{u}_{[4]} +\chi^y_{[1,1]}+\chi ^{a}_{[2]}  \chi^y_{[3,0]}+\chi ^{b}_{[2]}  \chi^y_{[3,0]}+\chi^y_{[3,0]}
\nonumber\\&\hspace{5cm}
+\chi ^{u}_{[2]}  (\chi ^{a}_{[2]} +\chi ^{b}_{[2]} -\chi^y_{[1,1]}+\chi^y_{[3,0]}-1)+2)q^8
\nonumber\\&
+(\chi ^{a}_{[1]}  \chi ^{b}_{[1]}  \chi ^{u}_{[3]}  \chi^y_{[2,0]}-\chi ^{a}_{[1]}  \chi ^{b}_{[1]}  \chi ^{u}_{[1]}  \chi^y_{[0,1]}) q^{\frac{26}{3}}
\nonumber\\&
+(2\chi ^{a}_{[2]}  \chi^y_{[1,0]}+\chi ^{a}_{[4]}  \chi^y_{[1,0]}+2 \chi ^{a}_{[2]}  \chi ^{b}_{[2]}  \chi^y_{[1,0]}+2 \chi ^{b}_{[2]}  \chi^y_{[1,0]}+\chi ^{b}_{[4]}  \chi^y_{[1,0]}+2 \chi ^{u}_{[4]}  \chi^y_{[1,0]}
\nonumber\\&\quad
+2 \chi^y_{[1,0]}+\chi^y_{[2,1]}+\chi ^{a}_{[2]}  \chi^y_{[4,0]}+\chi ^{b}_{[2]} 
   \chi^y_{[4,0]}+\chi^y_{[4,0]}
\nonumber\\&\quad
+\chi ^{u}_{[2]}  (2 \chi ^{a}_{[2]}  \chi^y_{[1,0]}+2 \chi ^{b}_{[2]}  \chi^y_{[1,0]}+2 \chi^y_{[1,0]}-\chi^y_{[2,1]}+\chi^y_{[4,0]})) q^{\frac{28}{3}}
\nonumber\\&
+(\chi ^{a}_{[1]}  \chi ^{b}_{[1]}  \chi^{u}_5-\chi ^{a}_{[1]}  \chi ^{b}_{[1]}  \chi ^{u}_{[1]}  \chi^y_{[1,1]}+\chi ^{u}_{[3]} 
   (2 \chi ^{a}_{[1]}  \chi ^{b}_{[1]} +\chi ^{a}_{[3]}  \chi ^{b}_{[1]} +\chi ^{a}_{[1]}  \chi^y_{[3,0]} \chi ^{b}_{[1]} +\chi ^{a}_{[1]}  \chi ^{b}_{[3]} ))
   q^{10}
\nonumber\\&
+(3 \chi ^{a}_{[2]}  \chi^y_{[0,1]}+\chi ^{a}_{[2]}  \chi ^{b}_{[2]}  \chi^y_{[0,1]}+3 \chi ^{b}_{[2]}  \chi^y_{[0,1]}-\chi^y_{[0,1]}+3
   \chi ^{a}_{[2]}  \chi^y_{[2,0]}+2 \chi ^{a}_{[4]}  \chi^y_{[2,0]}
\nonumber\\&\qquad
+3 \chi ^{a}_{[2]}  \chi ^{b}_{[2]}  \chi^y_{[2,0]}+3 \chi ^{b}_{[2]}  \chi^y_{[2,0]}+2
   \chi ^{b}_{[4]}  \chi^y_{[2,0]}+6 \chi^y_{[2,0]}+\chi ^{u}_{[4]}  (3 \chi^y_{[2,0]}-2 \chi^y_{[0,1]})
\nonumber\\&\qquad
+\chi^y_{[3,1]}+\chi ^{a}_{[2]}  \chi^y_{[5,0]}+\chi ^{b}_{[2]}  \chi^y_{[5,0]}+\chi^y_{[5,0]}+\chi ^{u}_{[2]}  (3 \chi^y_{[0,1]}+3 \chi ^{a}_{[2]}  \chi^y_{[2,0]}+3 \chi ^{b}_{[2]}  \chi^y_{[2,0]}
\nonumber\\&\qquad
+3 \chi^y_{[2,0]}-\chi^y_{[3,1]}+\chi^y_{[5,0]})) q^{\frac{32}{3}}
\nonumber\\&
+(2 \chi ^{a}_{[1]}  \chi ^{b}_{[1]}  \chi^{u}_{[5]} \chi^y_{[1,0]}+\chi ^{u}_{[1]} 
   (2 \chi ^{a}_{[1]}  \chi ^{b}_{[1]}  \chi^y_{[1,0]}-\chi ^{a}_{[1]}  \chi ^{b}_{[1]}  \chi^y_{[2,1]})+\chi ^{u}_{[3]}  (6 \chi ^{a}_{[1]}  \chi ^{b}_{[1]} 
   \chi^y_{[1,0]}
\nonumber\\&\qquad
+2 \chi ^{a}_{[3]}  \chi ^{b}_{[1]}  \chi^y_{[1,0]}+2 \chi ^{a}_{[1]}  \chi ^{b}_{[3]}  \chi^y_{[1,0]}+\chi ^{a}_{[1]}  \chi ^{b}_{[1]}  \chi^y_{[4,0]})) q^{\frac{34}{3}}
+\mathcal{O}\left(q^{12}\right).
%+(\chi ^{b}_{[4]}  \chi ^{a}_{[2]} +2 \chi ^{b}_{[2]}  \chi^y_{[1,1]} \chi ^{a}_{[2]} +5 \chi^y_{[1,1]} \chi ^{a}_{[2]} +4
%   \chi ^{b}_{[2]}  \chi^y_{[3,0]} \chi ^{a}_{[2]} +5 \chi^y_{[3,0]} \chi ^{a}_{[2]} +\chi^y_{[6,0]} \chi ^{a}_{[2]} 
%\nonumber\\&\qquad
%+3\chi ^{a}_{[2]} +\chi^{a}_6+\chi ^{a}_{[4]}  \chi ^{b}_{[2]} +3 \chi ^{b}_{[2]} +\chi^{b}_6+3 \chi^{u}_6+\chi ^{a}_{[4]}  \chi^y_{[1,1]}+5 \chi ^{b}_{[2]}  \chi^y_{[1,1]}+\chi ^{b}_{[4]}  \chi^y_{[1,1]}
%\nonumber\\&\qquad
%+3 \chi^y_{[1,1]}+2 \chi ^{a}_{[4]}  \chi^y_{[3,0]}+5 \chi ^{b}_{[2]}  \chi^y_{[3,0]}+2 \chi ^{b}_{[4]}  \chi^y_{[3,0]}+6 \chi^y_{[3,0]}+\chi ^{u}_{[4]}  (2 \chi ^{a}_{[2]} +2 \chi ^{b}_{[2]} 
%\nonumber\\&\qquad
%-2 \chi^y_{[1,1]}+3 \chi^y_{[3,0]}-2)+\chi^y_{[4,1]}+\chi ^{b}_{[2]}  \chi^y_{[6,0]}+\chi^y_{[6,0]}+\chi ^{u}_{[2]}  (\chi ^{b}_{[2]}  \chi ^{a}_{[2]} +4 \chi^y_{[3,0]}\chi ^{a}_{[2]} 
%\nonumber\\&\qquad
%-2 \chi ^{a}_{[2]} +\chi ^{a}_{[4]} -2 \chi ^{b}_{[2]} +\chi ^{b}_{[4]} +3 \chi^y_{[1,1]}+4 \chi ^{b}_{[2]}  \chi^y_{[3,0]}+5 \chi^y_{[3,0]}-\chi^y_{[4,1]}+\chi^y_{[6,0]})-1) q^{12}
\end{align}

%%%%%%%%%%%%%%%%%%%%%%%%%%%%%%%%%%%%%%%%%%%%%%%%%%%%
\subsubsection{$k=3$}
The global symmetry for $N=3$ and $k=3$ is
$G_{\rm flavor}=SU(3)_a\times SU(3)_b\times U(1)_F$.
The formula (\ref{idformula2}) gives
\begin{align}
\mathcal{I}_{N=3,k=3}^{\mathrm{grav}}
&=
1+(\chi^{a}_{[1,1]}+\chi^{b}_{[1,1]}+1) q^4+(\chi^{a}_{[1,1]} \chi^y_{[1,0]}+\chi^{b}_{[1,1]} \chi^y_{[1,0]}+2 \chi^y_{[1,0]})
   q^{\frac{16}{3}}
\nonumber\\&
   +(u^3\chi^{a}_{[1,0]} \chi^{b}_{[0,1]}+u^3+u^{-3}\chi^{a}_{[0,1]} \chi^{b}_{[1,0]}+u^{-3})
   q^6+(\chi^{a}_{[1,1]} \chi^y_{[2,0]}+\chi^{b}_{[1,1]} \chi^y_{[2,0]}+2 \chi^y_{[2,0]}) q^{\frac{20}{3}}
\nonumber\\&
   +(u^3\chi^{a}_{[1,0]}\chi^{b}_{[0,1]} \chi^y_{[1,0]}+u^3\chi^y_{[1,0]}
   +u^{-3}\chi^{a}_{[0,1]} \chi^{b}_{[1,0]} \chi^y_{[1,0]}+u^{-3}\chi^y_{[1,0]})q^{\frac{22}{3}}
\nonumber\\&
   +(\chi^{a}_{[1,1]}\chi^{b}_{[1,1]}+\chi^{a}_{[1,1]}\chi^y_{[3,0]}+2 \chi^{a}_{[1,1]}
   +\chi^{a}_{[2,2]}+\chi^{a}_{[1,0]}\chi^{b}_{[0,1]}
   \nonumber\\&\qquad
   +\chi^{a}_{[0,1]} \chi^{b}_{[1,0]}+2 \chi^{b}_{[1,1]}+\chi^{b}_{[2,2]}
   +\chi^{b}_{[1,1]} \chi^y_{[3,0]}+2 \chi^y_{[3,0]}+2) q^8
\nonumber\\&
   +(- u^3\chi^y_{[0,1]}+u^3\chi^{a}_{[1,0]} \chi^{b}_{[0,1]} \chi^y_{[2,0]}+u^3\chi^y_{[2,0]}
   -u^{-3}\chi^y_{[0,1]}+u^{-3}\chi^{a}_{[0,1]}\chi^{b}_{[1,0]} \chi^y_{[2,0]}+u^{-3}\chi^y_{[2,0]}) q^{\frac{26}{3}}
\nonumber\\&
   +(\chi^{a}_{[0,3]} \chi^y_{[1,0]}+5 \chi^{a}_{[1,1]} \chi^y_{[1,0]}+\chi^{a}_{[2,2]} \chi^y_{[1,0]}
   +\chi^{a}_{[3,0]} \chi^y_{[1,0]}+\chi^{a}_{[1,0]} \chi^{b}_{[0,1]} \chi^y_{[1,0]}+\chi^{b}_{[0,3]} \chi^y_{[1,0]}
   \nonumber\\&\qquad
   +\chi^{a}_{[0,1]} \chi^{b}_{[1,0]} \chi^y_{[1,0]}+2 \chi^{a}_{[1,1]} \chi^{b}_{[1,1]} \chi^y_{[1,0]}
   +5 \chi^{b}_{[1,1]} \chi^y_{[1,0]}+\chi^{b}_{[2,2]} \chi^y_{[1,0]}+\chi^{b}_{[3,0]} \chi^y_{[1,0]}
   \nonumber\\&\qquad
   +4 \chi^y_{[1,0]}+\chi^{a}_{[1,1]} \chi^y_{[4,0]}+\chi^{b}_{[1,1]} \chi^y_{[4,0]}+2 \chi^y_{[4,0]}) q^{\frac{28}{3}}
 \nonumber\\&
   +(u^3\chi^{a}_{[1,1]}+u^3\chi^{a}_{[0,2]} \chi^{b}_{[0,1]}+2 u^3 \chi^{a}_{[1,0]}\chi^{b}_{[0,1]}
   +u^3\chi^{a}_{[2,1]} \chi^{b}_{[0,1]}+u^3\chi^{a}_{[0,1]} \chi^{b}_{[1,0]}+u^3\chi^{b}_{[1,1]}
   \nonumber\\&\qquad
   +u^3\chi^{a}_{[1,0]} \chi^{b}_{[1,2]}+u^3\chi^{a}_{[1,0]} \chi^{b}_{[2,0]}-u^3\chi^y_{[1,1]}
   +u^3\chi^{a}_{[1,0]} \chi^{b}_{[0,1]}\chi^y_{[3,0]}+u^3\chi^y_{[3,0]}+u^3
   \nonumber\\&\qquad
   +u^{-3}\chi^{a}_{[1,1]}+u^{-3}\chi^{a}_{[1,0]} \chi^{b}_{[0,1]}+u^{-3}\chi^{a}_{[0,1]} \chi^{b}_{[0,2]}+2u^{-3}
   \chi^{a}_{[0,1]} \chi^{b}_{[1,0]}+u^{-3}\chi^{a}_{[1,2]} \chi^{b}_{[1,0]}
   \nonumber\\&\qquad
   +u^{-3}\chi^{a}_{[2,0]} \chi^{b}_{[1,0]}+u^{-3}\chi^{b}_{[1,1]}+u^{-3}\chi^{a}_{[0,1]}
   \chi^{b}_{[2,1]}-u^{-3}\chi^y_{[1,1]}+u^{-3}\chi^{a}_{[0,1]} \chi^{b}_{[1,0]} \chi^y_{[3,0]}
   \nonumber\\&\qquad
   +u^{-3}\chi^y_{[3,0]}+u^{-3}) q^{10}
\nonumber\\&
   +(\chi^{a}_{[0,3]}\chi^y_{[0,1]}+3 \chi^{a}_{[1,1]} \chi^y_{[0,1]}+\chi^{a}_{[3,0]} \chi^y_{[0,1]}
   -\chi^{a}_{[1,0]} \chi^{b}_{[0,1]} \chi^y_{[0,1]}
   +\chi^{b}_{[0,3]} \chi^y_{[0,1]}-\chi^{a}_{[0,1]} \chi^{b}_{[1,0]} \chi^y_{[0,1]}
   \nonumber\\&\qquad
   +\chi^{a}_{[1,1]} \chi^{b}_{[1,1]} \chi^y_{[0,1]}+3\chi^{b}_{[1,1]} \chi^y_{[0,1]}+\chi^{b}_{[3,0]} \chi^y_{[0,1]}
   +\chi^y_{[0,1]}+\chi^{a}_{[0,3]} \chi^y_{[2,0]}+8 \chi^{a}_{[1,1]} \chi^y_{[2,0]}
   \nonumber\\&\qquad
   +2\chi^{a}_{[2,2]} \chi^y_{[2,0]}+\chi^{a}_{[3,0]} \chi^y_{[2,0]}+\chi^{a}_{[1,0]} \chi^{b}_{[0,1]} \chi^y_{[2,0]}
   +\chi^{b}_{[0,3]} \chi^y_{[2,0]}+\chi^{a}_{[0,1]} \chi^{b}_{[1,0]} \chi^y_{[2,0]}
   \nonumber\\&\qquad
   +3 \chi^{a}_{[1,1]} \chi^{b}_{[1,1]} \chi^y_{[2,0]}
   +8 \chi^{b}_{[1,1]} \chi^y_{[2,0]}+2\chi^{b}_{[2,2]} \chi^y_{[2,0]}+\chi^{b}_{[3,0]} \chi^y_{[2,0]}
   +9 \chi^y_{[2,0]} +\chi^{a}_{[1,1]} \chi^y_{[5,0]}
   \nonumber\\&\qquad
  +\chi^{b}_{[1,1]} \chi^y_{[5,0]}+2\chi^y_{[5,0]}) q^{\frac{32}{3}}
\nonumber\\&
   +(2u^3 \chi^{a}_{[1,1]} \chi^y_{[1,0]}+2u^3 \chi^{a}_{[0,2]} \chi^{b}_{[0,1]} \chi^y_{[1,0]}
   +6u^3\chi^{a}_{[1,0]} \chi^{b}_{[0,1]} \chi^y_{[1,0]}
   +2 u^3\chi^{a}_{[2,1]} \chi^{b}_{[0,1]} \chi^y_{[1,0]}
   \nonumber\\&\qquad
   +u^3\chi^{a}_{[0,1]}\chi^{b}_{[1,0]} \chi^y_{[1,0]}
   +2 u^3\chi^{b}_{[1,1]} \chi^y_{[1,0]}+2 u^3\chi^{a}_{[1,0]} \chi^{b}_{[1,2]} \chi^y_{[1,0]}
   +2 u^3\chi^{a}_{[1,0]} \chi^{b}_{[2,0]} \chi^y_{[1,0]}
   \nonumber\\&\qquad
   +4 u^3\chi^y_{[1,0]}-u^3\chi^y_{[2,1]}+u^3\chi^{a}_{[1,0]} \chi^{b}_{[0,1]} \chi^y_{[4,0]}
   +u^3\chi^y_{[4,0]}+2 u^{-3}\chi^{a}_{[1,1]} \chi^y_{[1,0]}
   \nonumber\\&\qquad
   +u^{-3}\chi^{a}_{[1,0]} \chi^{b}_{[0,1]} \chi^y_{[1,0]}+2 u^{-3}\chi^{a}_{[0,1]} \chi^{b}_{[0,2]}
   \chi^y_{[1,0]}+6 u^{-3}\chi^{a}_{[0,1]} \chi^{b}_{[1,0]} \chi^y_{[1,0]}+2 u^{-3}\chi^{a}_{[1,2]} \chi^{b}_{[1,0]} \chi^y_{[1,0]}
   \nonumber\\&\qquad
   +2 u^{-3}\chi^{a}_{[2,0]}\chi^{b}_{[1,0]} \chi^y_{[1,0]}+2 u^{-3}\chi^{b}_{[1,1]} \chi^y_{[1,0]}
   +2 u^{-3}\chi^{a}_{[0,1]} \chi^{b}_{[2,1]} \chi^y_{[1,0]}+4 u^{-3}\chi^y_{[1,0]}
   \nonumber\\&\qquad
   -u^{-3}\chi^y_{[2,1]}+u^{-3}\chi^{a}_{[0,1]} \chi^{b}_{[1,0]} \chi^y_{[4,0]}+u^{-3}\chi^y_{[4,0]})q^{\frac{34}{3}}
+\mathcal{O}\left(q^{12}\right).
\end{align}

%%%%%%%%%%%%%%%%%%%%%%%%%%%%%%%%%%%%%%%%%%%%%%%%%%%%%%%%%%%%%%%%
\section{Conclusions and discussions}\label{disc.sec}
In this paper we calculated the superconformal index of
${\cal N}=(1,0)$ theories realized on $N$ M5-branes at the $\CC^2/\ZZ_k$ singularity.
We used the holographic description of the theories, M-theory on $AdS_7\times S^4/\ZZ_k$.
For small $k$ and $N$ we confirmed that the indices are consistent with the expected
flavor symmetries in Table \ref{flavor.tbl}.
Namely, each of the index is expanded in terms of characters of the flavor symmetry.
To see the symmetry enhancement at $N=2$ it is crucial to include the
finite $N$ corrections due to wrapped M2-branes.
We use the formula (\ref{idformula}) including the single-wrapping contributions.

In fact for the $N=1$ case the 6d index was reproduced from 5d abelian quiver theories in \cite{Benvenuti:2016dcs}.
It is nice if one can generalize their results and compare directly with our results for $N\geq 2$.

There are many ways of extension.
We consider only the simple class of ${\cal N}=(1,0)$ theories realized on a D6-NS5 system.
It is known that this is generalized by introducong D8-branes \cite{Hanany:1997sa,Brunner:1997gf,Hanany:1997gh}.
Because the corresponding supergravity solutions are known \cite{Apruzzi:2013yva,Gaiotto:2014lca,Apruzzi:2015wna}
it would be possible to apply our method to these theories.
Inclusion of orientifold planes and M9-planes \cite{Brunner:1997gk,Berkooz:1998bx,Apruzzi:2017nck}
may also be interesting.

In this paper we focus only on the single-wrapping contributions.
This is because we have not yet understood how to determine the
integration contours in the gauge integrals that we need to perform
to calculate the contribution of multiple branes.
In the case of $(2,0)$ theory
we can define Schur-like index
and it was found in \cite{Arai:2020uwd}
that by adopting an appropriate pole selection rule
we can reproduce the known Schur-like index.
Although we cannot define the Schur-like index
for ${\cal N}=(1,0)$ theory, unfortunately,
this strongly suggests that
we can reproduce the all order index by including multiple-brane contributions.

%%%%%%%%%%%%%%%%%%%%%%%%%%%%%%%%%%%%%%%%%%%%%%%%%%%%%%%%%%%%%%%%
\section*{Acknowledgments}
We would like to thank R. Arai and D. Yokoyama for wonderful discussions and useful comments.

%%%%%%%%%%%%%%%%%%%%%%%%%%%%%%%%%%%%%%%%%%%%%%%%%%%%


\begin{thebibliography}{99}



%\cite{Brunner:1997gk}
\bibitem{Brunner:1997gk}
I.~Brunner and A.~Karch,
``Branes and six-dimensional fixed points,''
Phys. Lett. B \textbf{409}, 109-116 (1997)
doi:10.1016/S0370-2693(97)00935-0
[arXiv:hep-th/9705022 [hep-th]].
%89 citations counted in INSPIRE as of 20 Mar 2021

%\cite{Hanany:1997sa}
\bibitem{Hanany:1997sa}
A.~Hanany and A.~Zaffaroni,
``Chiral symmetry from type IIA branes,''
Nucl. Phys. B \textbf{509}, 145-168 (1998)
doi:10.1016/S0550-3213(97)00595-6
[arXiv:hep-th/9706047 [hep-th]].
%66 citations counted in INSPIRE as of 20 Mar 2021

%\cite{Brunner:1997gf}
\bibitem{Brunner:1997gf}
I.~Brunner and A.~Karch,
``Branes at orbifolds versus Hanany Witten in six-dimensions,''
JHEP \textbf{03}, 003 (1998)
doi:10.1088/1126-6708/1998/03/003
[arXiv:hep-th/9712143 [hep-th]].
%153 citations counted in INSPIRE as of 20 Mar 2021

%\cite{Hanany:1997gh}
\bibitem{Hanany:1997gh}
A.~Hanany and A.~Zaffaroni,
``Branes and six-dimensional supersymmetric theories,''
Nucl. Phys. B \textbf{529}, 180-206 (1998)
doi:10.1016/S0550-3213(98)00355-1
[arXiv:hep-th/9712145 [hep-th]].
%178 citations counted in INSPIRE as of 20 Mar 2021

%\cite{Ohmori:2015pia}
\bibitem{Ohmori:2015pia}
K.~Ohmori, H.~Shimizu, Y.~Tachikawa and K.~Yonekura,
``6d $\mathcal{N}=\left(1,\;0\right) $ theories on S$^{1}$ /T$^{2}$ and class S theories: part II,''
JHEP \textbf{12}, 131 (2015)
doi:10.1007/JHEP12(2015)131
[arXiv:1508.00915 [hep-th]].
%82 citations counted in INSPIRE as of 20 Mar 2021

%\cite{Hanany:2018vph}
\bibitem{Hanany:2018vph}
A.~Hanany and G.~Zafrir,
``Discrete Gauging in Six Dimensions,''
JHEP \textbf{07}, 168 (2018)
doi:10.1007/JHEP07(2018)168
[arXiv:1804.08857 [hep-th]].
%31 citations counted in INSPIRE as of 20 Mar 2021

%\cite{Bergman:2020bvi}
\bibitem{Bergman:2020bvi}
O.~Bergman, M.~Fazzi, D.~Rodr\'\i{}guez-G\'omez and A.~Tomasiello,
``Charges and holography in 6d (1,0) theories,''
JHEP \textbf{05}, 138 (2020)
doi:10.1007/JHEP05(2020)138
[arXiv:2002.04036 [hep-th]].
%7 citations counted in INSPIRE as of 20 Mar 2021

%\cite{Apruzzi:2020eqi}
\bibitem{Apruzzi:2020eqi}
F.~Apruzzi, M.~Fazzi, J.~J.~Heckman, T.~Rudelius and H.~Y.~Zhang,
``General prescription for global $U(1)$\textquoteright{}s in 6D SCFTs,''
Phys. Rev. D \textbf{101}, no.8, 086023 (2020)
doi:10.1103/PhysRevD.101.086023
[arXiv:2001.10549 [hep-th]].
%10 citations counted in INSPIRE as of 15 Apr 2021

%\cite{Bertolini:2015bwa}
\bibitem{Bertolini:2015bwa}
M.~Bertolini, P.~R.~Merkx and D.~R.~Morrison,
``On the global symmetries of 6D superconformal field theories,''
JHEP \textbf{07}, 005 (2016)
doi:10.1007/JHEP07(2016)005
[arXiv:1510.08056 [hep-th]].
%28 citations counted in INSPIRE as of 28 Apr 2021

%\cite{Merkx:2017jey}
\bibitem{Merkx:2017jey}
P.~R.~Merkx,
``Classifying Global Symmetries of 6D SCFTs,''
JHEP \textbf{03}, 163 (2018)
doi:10.1007/JHEP03(2018)163
[arXiv:1711.05155 [hep-th]].
%10 citations counted in INSPIRE as of 28 Apr 2021

%\cite{Bhardwaj:2020ruf}
\bibitem{Bhardwaj:2020ruf}
L.~Bhardwaj,
``Flavor Symmetry of 5d SCFTs, Part 1: General Setup,''
[arXiv:2010.13230 [hep-th]].
%4 citations counted in INSPIRE as of 28 Apr 2021

%\cite{Arai:2020uwd}
\bibitem{Arai:2020uwd}
R.~Arai, S.~Fujiwara, Y.~Imamura, T.~Mori and D.~Yokoyama,
``Finite-$N$ corrections to the M-brane indices,''
JHEP \textbf{11}, 093 (2020)
doi:10.1007/JHEP11(2020)093
[arXiv:2007.05213 [hep-th]].
%1 citations counted in INSPIRE as of 20 Mar 2021

%\cite{Arai:2019xmp}
\bibitem{Arai:2019xmp}
R.~Arai and Y.~Imamura,
``Finite $N$ Corrections to the Superconformal Index of S-fold Theories,''
PTEP \textbf{2019}, no.8, 083B04 (2019)
doi:10.1093/ptep/ptz088
[arXiv:1904.09776 [hep-th]].
%7 citations counted in INSPIRE as of 20 Mar 2021

%\cite{Arai:2019wgv}
\bibitem{Arai:2019wgv}
R.~Arai, S.~Fujiwara, Y.~Imamura and T.~Mori,
``Finite $N$ corrections to the superconformal index of orbifold quiver gauge theories,''
JHEP \textbf{10}, 243 (2019)
doi:10.1007/JHEP10(2019)243
[arXiv:1907.05660 [hep-th]].
%3 citations counted in INSPIRE as of 20 Mar 2021

%\cite{Arai:2019aou}
\bibitem{Arai:2019aou}
R.~Arai, S.~Fujiwara, Y.~Imamura and T.~Mori,
``Finite $N$ corrections to the superconformal index of toric quiver gauge theories,''
PTEP \textbf{2020}, no.4, 043B09 (2020)
doi:10.1093/ptep/ptaa023
[arXiv:1911.10794 [hep-th]].
%2 citations counted in INSPIRE as of 20 Mar 2021

%\cite{Arai:2020qaj}
\bibitem{Arai:2020qaj}
R.~Arai, S.~Fujiwara, Y.~Imamura and T.~Mori,
``Schur index of the ${\cal N}=4$ $U(N)$ supersymmetric Yang-Mills theory via the AdS/CFT correspondence,''
Phys. Rev. D \textbf{101}, no.8, 086017 (2020)
doi:10.1103/PhysRevD.101.086017
[arXiv:2001.11667 [hep-th]].
%1 citations counted in INSPIRE as of 20 Mar 2021

%\cite{Bhattacharya:2008zy}
\bibitem{Bhattacharya:2008zy}
J.~Bhattacharya, S.~Bhattacharyya, S.~Minwalla and S.~Raju,
``Indices for Superconformal Field Theories in 3,5 and 6 Dimensions,''
JHEP \textbf{02}, 064 (2008)
doi:10.1088/1126-6708/2008/02/064
[arXiv:0801.1435 [hep-th]].
%191 citations counted in INSPIRE as of 22 Mar 2021

%\cite{Ahn:1998pb}
\bibitem{Ahn:1998pb}
C.~h.~Ahn, K.~Oh and R.~Tatar,
``Orbifolds of AdS(7) x S**4 and six-dimensional (0,1) SCFT,''
Phys. Lett. B \textbf{442}, 109-116 (1998)
doi:10.1016/S0370-2693(98)01276-3
[arXiv:hep-th/9804093 [hep-th]].
%27 citations counted in INSPIRE as of 20 Mar 2021


%\cite{Mikhailov:2000ya}
\bibitem{Mikhailov:2000ya}
A.~Mikhailov,
``Giant gravitons from holomorphic surfaces,''
JHEP \textbf{11}, 027 (2000)
doi:10.1088/1126-6708/2000/11/027
[arXiv:hep-th/0010206 [hep-th]].
%140 citations counted in INSPIRE as of 20 Mar 2021

%\cite{Bhattacharyya:2007sa}
\bibitem{Bhattacharyya:2007sa}
S.~Bhattacharyya and S.~Minwalla,
``Supersymmetric states in M5/M2 CFTs,''
JHEP \textbf{12}, 004 (2007)
doi:10.1088/1126-6708/2007/12/004
[arXiv:hep-th/0702069 [hep-th]].
%38 citations counted in INSPIRE as of 20 Mar 2021

%\cite{Bah:2017gph}
\bibitem{Bah:2017gph}
I.~Bah, A.~Hanany, K.~Maruyoshi, S.~S.~Razamat, Y.~Tachikawa and G.~Zafrir,
``4d $ \mathcal{N}=1 $ from 6d $ \mathcal{N}=\left(1,0\right) $ on a torus with fluxes,''
JHEP \textbf{06}, 022 (2017)
doi:10.1007/JHEP06(2017)022
[arXiv:1702.04740 [hep-th]].
%44 citations counted in INSPIRE as of 20 Mar 2021

%\cite{Benvenuti:2016dcs}
\bibitem{Benvenuti:2016dcs}
S.~Benvenuti, G.~Bonelli, M.~Ronzani and A.~Tanzini,
``Symmetry enhancements via 5d instantons, $ q\mathcal{W} $ -algebrae and (1, 0) superconformal index,''
JHEP \textbf{09}, 053 (2016)
doi:10.1007/JHEP09(2016)053
[arXiv:1606.03036 [hep-th]].
%14 citations counted in INSPIRE as of 15 Sep 2020

%\cite{Apruzzi:2013yva}
\bibitem{Apruzzi:2013yva}
F.~Apruzzi, M.~Fazzi, D.~Rosa and A.~Tomasiello,
``All AdS$_7$ solutions of type II supergravity,''
JHEP \textbf{04}, 064 (2014)
doi:10.1007/JHEP04(2014)064
[arXiv:1309.2949 [hep-th]].
%147 citations counted in INSPIRE as of 20 Mar 2021

%\cite{Gaiotto:2014lca}
\bibitem{Gaiotto:2014lca}
D.~Gaiotto and A.~Tomasiello,
``Holography for (1,0) theories in six dimensions,''
JHEP \textbf{12}, 003 (2014)
doi:10.1007/JHEP12(2014)003
[arXiv:1404.0711 [hep-th]].
%115 citations counted in INSPIRE as of 20 Mar 2021

%\cite{Apruzzi:2015wna}
\bibitem{Apruzzi:2015wna}
F.~Apruzzi, M.~Fazzi, A.~Passias, A.~Rota and A.~Tomasiello,
``Six-Dimensional Superconformal Theories and their Compactifications from Type IIA Supergravity,''
Phys. Rev. Lett. \textbf{115}, no.6, 061601 (2015)
doi:10.1103/PhysRevLett.115.061601
[arXiv:1502.06616 [hep-th]].
%58 citations counted in INSPIRE as of 20 Mar 2021

%\cite{Berkooz:1998bx}
\bibitem{Berkooz:1998bx}
M.~Berkooz,
``A Supergravity dual of a (1,0) field theory in six-dimensions,''
Phys. Lett. B \textbf{437}, 315-317 (1998)
doi:10.1016/S0370-2693(98)00913-7
[arXiv:hep-th/9802195 [hep-th]].
%56 citations counted in INSPIRE as of 20 Mar 2021

%\cite{Apruzzi:2017nck}
\bibitem{Apruzzi:2017nck}
F.~Apruzzi and M.~Fazzi,
``AdS$_{7}$/CFT$_{6}$ with orientifolds,''
JHEP \textbf{01}, 124 (2018)
doi:10.1007/JHEP01(2018)124
[arXiv:1712.03235 [hep-th]].
%35 citations counted in INSPIRE as of 15 Apr 2021


\end{thebibliography}
\end{document}